\documentclass[final,5p,times,twocolumn]{elsarticle}

\usepackage{epsfig}

\usepackage{amssymb}

\journal{Nuclear Physics B}
\begin{document}

\begin{frontmatter}



\title{IceCube's In-Ice Radio Extension: Status and Results}


\author[uw]{H. Landsman} \author{for the IceCube Collaboration\fnref{fn1}}
\author[uw]{E.~Cheng}
\author[uw]{E.~Kulcyk}
\author[uw]{A.W.~Laundrie}
\author[uh]{B.~Rotter}
\author[uh]{L.~Ruckman}
\author[uw]{P.W.~Sandstrom}
\author[uh]{and G.S.~Varner}
\address[uw]{Dept. of Physics, Univ. of Wisconsin, Madison, WI 53703, USA}
\address[uh]{Dept. of Physics and Astronomy, Univ. of Hawaii, Manoa, HI 96822, USA}
\fntext[fn1]{http://www.icecube.wisc.edu}
\begin{abstract}
In 2006-2010, several Radio Frequency (RF) detectors and calibration 
equipment were deployed as part of the IceCube array at depths between 
5 to 1400 meters in preparation for a future large scale GZK neutrino detector. 
IceCube's deep holes and well-established data handling system provide a 
unique opportunity for deep-ice RF detection studies at the South-Pole. 

We will present verification and calibration results as well as 
a status-review of ongoing analyses such as ice-properties, RF noise and
reconstruction algorithms.

\end{abstract}

\begin{keyword}

Neutrino astronomy \sep Radio frequency \sep RF \sep Ice \sep South Pole \sep GZK neutrinos \sep Neutrinos detection \sep Ice properties


\end{keyword}

\end{frontmatter}


\section{Introduction}
\label{Introduction}
Astrophysical high-energy neutrinos may carry valuable information about their origin, possibly point sources such as GRBs, AGNs, or SGRs, or high-energy cosmic rays produced in the GZK process; they may also offer the chance to investigate particle physics in a very high energy range unreachable by earthbound accelerators.
Designed to detect neutrinos with energies in the range of  $10^{2}-10^{10}\,\mbox{GeV}$, currently existing neutrino detectors, such as IceCube and Antares, are quite large, incorporating hundreds of photo-multiplier tubes, sensitive to photons in the optical wavelength range.  In order to survey the extremely high energy regime above $10^{10}\,\mbox{GeV}$, however, even larger detectors will be needed.
The concept of a GZK neutrino radio frequency detector buried in ice at shallow depths or deployed as a surface array was suggested more than 20 years ago \cite{Gusev:1984gh}. The RICE\cite{Kravchenko:2006qc} array and  the ANITA experiment\cite{Gorham:2010kv} are already taking advantage of the Askaryan effect and the massive volume of ice in Antarctica for neutrino detection. Future experiments include ARIANNA\cite{Gerhardt:2010js}, a surface array on the Ross Ice Shelf, and ARA\cite{ara}, an in-ice array at shallow depths near the South Pole.


Our unique access to IceCube's deep and wide holes have provided us with an opportunity for deploying radio frequency (RF) detectors in the deep Polar ice. These detectors use the communication and time calibration systems developed for IceCube and rely on the experience within the IceCube Collaboration for developing hardware and software and for building and deploying highly-sensitive equipment in the extreme South Pole environment as well as on radio technology expertise from the RICE and ANITA Collaborations.

\begin{figure}[htbp]
\centering 
\noindent
\includegraphics*[width=7cm]{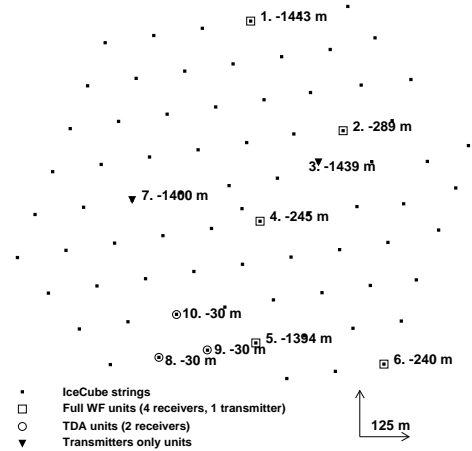}
\caption{Map of the IceCube radio detector deployments, plotted on top of the full IceCube array. Also shown the deployment depth of each unit relative to the surface.  The distance between adjacent strings is $125\,\mbox{m}$. A description of each unit is summarized in Table \ref{Deploy_Table}. }
\label{fig:AURAGeo}
\end{figure}

In the following sections, the design of the deployed detectors will be discussed as well as some of the on-going studies using these detectors. Figure  \ref{fig:AURAGeo} and Table \ref{Deploy_Table} summarizes the in-ice locations and depths of the units described below. 

\section{Hardware description}
IceCube's radio extension modules were installed during the Austral summers between 2006 and 2010 at depths of $5$ to $1400\,\mbox{m}$.  Each radio module was installed directly above IceCube's digital optical modules (DOMs) immediately following a successful installation of the string. The RF components were mechanically attached to the $\sim3\,\mbox{km}$-long IceCube main cable (a 5-cm-wide cable bundle for communication and power) which simply served as a mechanical support; some of the modules, however, were tapped into one of the auxiliary twisted-pairs within the main cable designed for specialized device operations, such as this one.  The main cable, being a massive conducting object, can shadow the RF antennas, as discussed later.

During the first two seasons five detectors capable of full waveform (WF) digitization were deployed. These so called  ``clusters'' consisted of a set of antennas and the central electronics module, and were built upon the experiences and technologies of ANITA, RICE, and IceCube. During the last season, a small array of envelope-detectors for transient detection study and additional calibration hardware were deployed at shallow depth.
\subsection{Season I (2006-2007) and Season II (2008-2009): Full-digitization detectors }
In the 2006-2007 season, two clusters were deployed at two separate depths of $\sim300\,\mbox{m}$ and $\sim1400\,\mbox{m}$. Each cluster consists of a central sphere containing electronics (DRM:Digital Radio Module), a transmitting antenna for calibration (ACU: antenna calibration unit), and four receiver antennas. Each receiver channel consists of a `Fat Dipole' antenna (with the gain centered at $\sim400\,\mbox{MHz}$ in air and $\sim250\,\mbox{MHz}$ in ice),  and a set of front-end electronics (housed in a metal tube), including filters and amplifiers necessary for proper channel operation. Specifically, included in the front-end electronics are a $450$ MHz notch filter for rejecting constant interference from the South Pole communication channel, a $200\,\mbox{MHz}$ high-pass filter, and a $\sim50\,\mbox{dB}$ low-noise amplifier (LNA). An additional $\sim20\,\mbox{dB}$ amplification is done at a later stage within the DRM for a total amplification of $\sim70\,\mbox{dB}$. The antennas are equally spaced over 40 m along the IceCube in-ice cable. A schematic of the 2006-2007  cluster is shown in Fig. \ref{fig:cluster-r}. Triggering and digitization is done in a similar way to ANITA \cite{labrador}, namely, by splitting the signal from individual channel into four frequency bands and requiring threshold crossing in a programmed number of frequency bands and channels. The sampling speed is two giga-samples per second into a 256 ns-deep buffer. The digitized data are sent to the surface using the IceCube cable system. During the 2006-2007 season, a third cluster with only a transmission channel was also deployed.
A detailed description of the electronics installed inside the DRM can be found elsewhere \cite{Landsman:2008rx}.

During the Austral summer of 2008-2009, three additional radio clusters were deployed. These clusters are based on the same design as above with a few minor modifications, including an addition of two low-frequency channels sensitive down to 100 MHz, an ACU with a higher transmission power, and a shorter spacing between antennas. The trigger logic board was also modified to act on three frequency bands (instead of four) plus the full-frequency band.

\subsection{Season III(2009-2010): Transient detectors and calibration antennas}The idea of using an array of simple transient sensors to image the unique spatiotemporal signature of neutrino interactions in Antarctic ice was proposed by Gusev and Zheleznykh nearly thirty years ago \cite{Gusev:1984gh}. In this type of detector system, the pattern of coincident hits among a large number of sensors provides event confirmation, indication of direction and energy, and information for rejecting sporadic noise on the basis of time-of-arrival and amplitude.
  
\begin{figure}[htbp]
  \begin{minipage}[b]{0.38\linewidth}
    \centering
    \includegraphics[width=\linewidth]{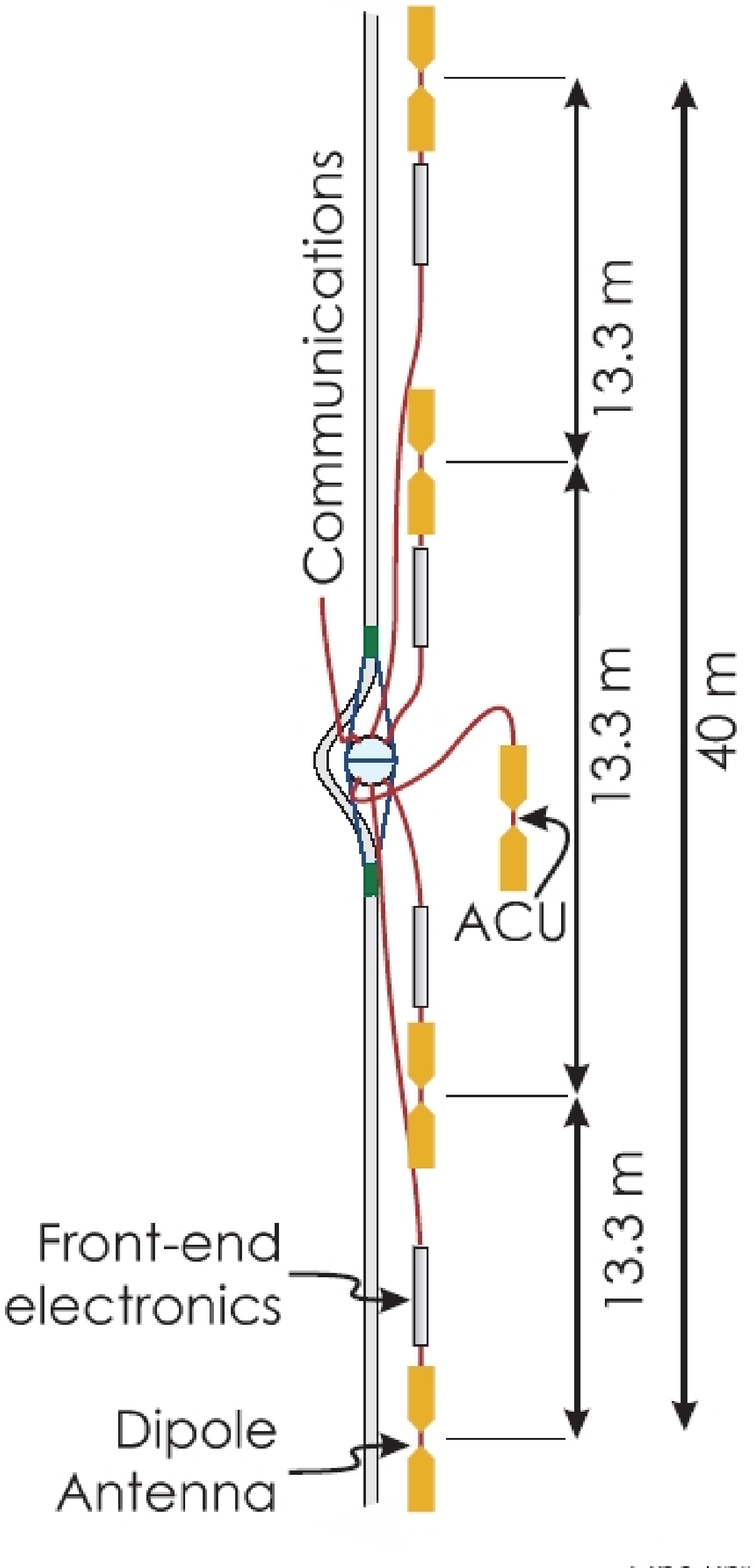}
    \caption{The radio cluster, consisting of the DRM (Digital Radio Module) and 5 antennas (4 receivers and a transmitter).}
    \label{fig:cluster-r}
  \end{minipage}
  \hspace{.4cm}
  \begin{minipage}[b]{0.35\linewidth}
    \centering
    \includegraphics[width=\linewidth]{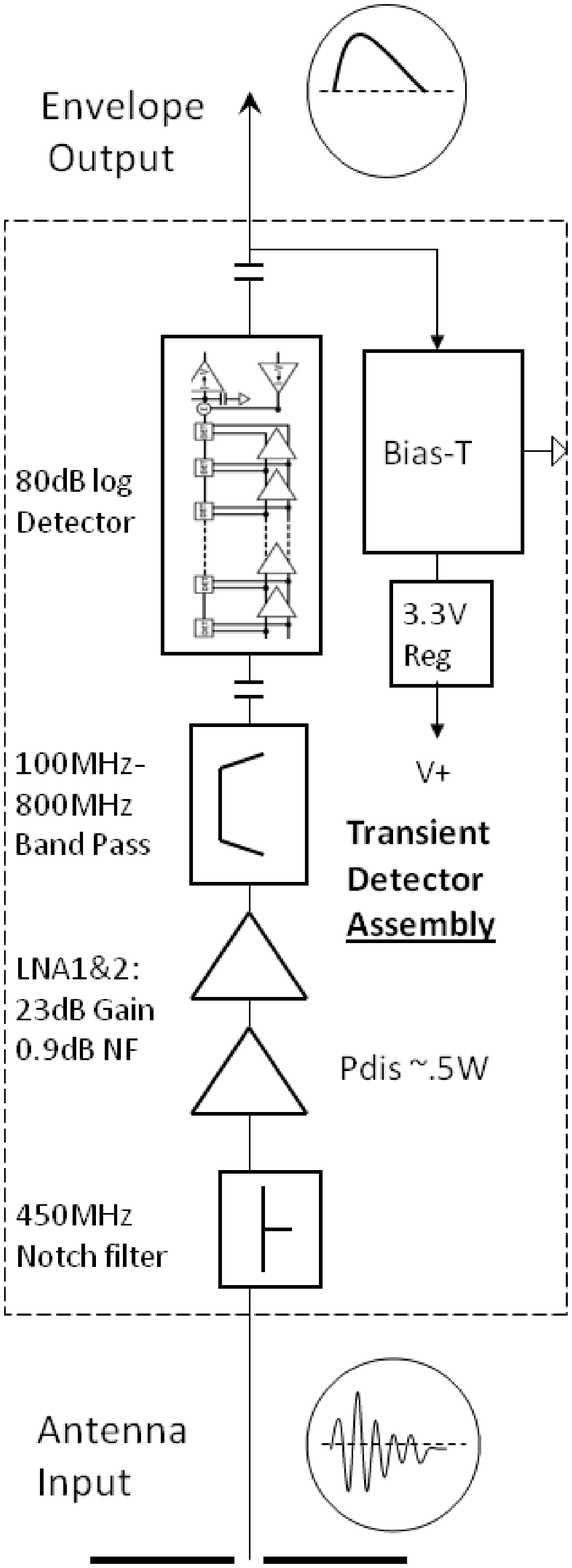}
    \caption{Block diagram of the Transient Detector Assembly.}
    \label{fig:TDABlock}
  \end{minipage}
\end{figure}

Six units of transient-prototype-detector were deployed in the 2009-2010 season. Each unit consists of a discone wide-band omni-directional antenna feeding into a Transient Detector Assembly (TDA), an exploratory device, whose block diagram is shown in Fig. \ref{fig:TDABlock}. Each unit is read out using an unmodified control Mother-Board (MB) developed for the IceCube DOM \cite{IceCube}. The IceCube cable and calibration system also facilitates timing calibration and data handling. The units were deployed in pairs above three IceCube strings, with one unit at $z=-5\,\mbox{m}$ and the other at $z=-35\,\mbox{m}$. The Local Coincidence (LC) capabilities of the IceCube MB are also exploited; when LC is enabled, each TDA pair reads out data only if both of the pair are triggered in some adjustable time window. The rise-time of the output pulse from these units is on the order of 10-20 ns and is largely independent of amplitude. 

Also deployed during the same season are a set of azimuthal-symmetric antennas for the attenuation length measurements  mounted symmetrically around the support cable rather than adjacent to it.

\begin{table*}
\begin{center}
\caption{Lateral coordinates and depth of the RF hardware with additional information including number of antennas (Tx: transmitting; Rx: receiving) and detector type. Coordinates are relative to the center of the IceCube array at surface (See Fig \ref{fig:AURAGeo})} \label{Deploy_Table}
\begin{tabular}{|c|l|c|c|c|l|c|l|}
\hline 
num&Name& Season deployed & Tx & Rx & Location (x,y,z) [m]& IceCube hole  & Type \\
\hline
1 &Doris & I & 1 & 4 & $(22,509,-1443)$ & 78 & Full digitization \\
2 &Sophie& I & 1 & 4 & $(257,211,-289)$ & 57 &Full digitization \\
3 &Tx2006& I & 1 & 0 & $(195,126,-1439)$&47 &  Transmitter only\\
4 &Sally & II &1 & 3+1 low freq. & $(46,-34,-245)$& 36& Full digitization\\
5 &Danielle & II &1 & 3+1 low freq & $(35,-365,-1394)$&10 & Full digitization \\
6 &Susan& II &1 & 4& $(361,-423,-240)$ &6& Full digitization\\
7 &Tx2010& III &2 &0 & $(-280,23,-1400)$ &43 &Transmitter only \\
8 &Top8 and Bottom8 &  III &0 &2 &$(-210,-404,-5/-35)$ &8 &Transient detection \\
9 &Top9 and Bottom9 &III&0 &2 &$(-88,-384,-5/-35)$ &9 &Transient detection \\
10& Top16 and Bottom16 &III&0 &2 &$(-166,-288,-5/-35)$ &16 &Transient detection \\
\hline
\end{tabular}
\end{center}
\end{table*}

\section{South Pole ice and ray tracing}
The properties of ice at radio frequencies determine the feasibility and design of future GZK neutrino detectors. Specifically, the attenuation length affects the spacing between channels and the effective detector volume, whereas the index of refraction determines the reconstruction capabilities and simulation quality.

The existing in-ice RF attenuation data from the South Pole were obtained by sending signals down into the ice using a surface transmitter and recording the signals reflected on the bedrock below. This measurement provided an average RF attenuation over the round-trip, weighted by the temperature profile along the path \cite{ice-atten}.

The latest index of refraction measurement at the South Pole reported in \cite{ice-n} combines results using the RICE array down to 150 m and ice cores down to 240 m. The model used is of the form  $ n(z)=n_{deep} + (n_{shallow} - n_{deep})e^{n_c \cdot z} $, where $n_c\approx -0.0132\,{\mbox{m}}^{-1}$, $n_{deep}\approx 1.78$, and  $n_{shallow} \approx 1.35$. The changing index of refraction causes rays to curve in the ice layers, especially in the soft ice layers on top of the glacier (firn), and decreases the angular acceptance of shallow-deployed detectors by causing total reflection of rays propagating between the layers. Since there is no analytical solution to the ray-tracing problem, detailed simulation is needed for accurate ray tracing; this, however, requires extensive computation. We have been able to decrease the computation time by an order of magnitude using a semi-analytical method.

Uncertainties in index of refraction lead to uncertainties of the order of a few ns in the time difference measured between two receivers a few meters apart (geometry-dependent); therefore, a precise knowledge of index of refraction is necessary for a sub-ns-resolution detector.

\begin{figure}[htbp]
\centering
\noindent
\includegraphics*[width=6.3cm]{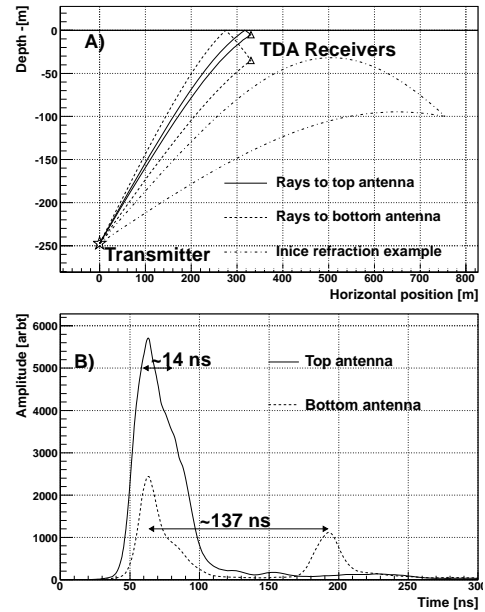}
\caption{A. Some examples of Ray Tracing solutions showing the direct- and refracted-rays solutions. The depths and separation between the transmitter and the receivers in this figure correspond to the geometry described in the text of ``Sally'''s transmitter and the transient detectors in hole 16. B. Average WF collected on the transient detectors in hole 16 for the top (solid) and bottom (dashed) antennas. The time delay between the direct and the refracted ray detected by the bottom antenna is consistent with simulations (137 ns).}\label{fig:RayTrace_sally_satra16}
\end{figure}

When looking at possible paths connecting a transmitter and a receiver both in ice there will either be no solution or two solutions (direct ray and refracted ray) to the ray-tracing problem. The refraction may take place at the ice-air boundary on the surface, but also within ice, as illustrated in Fig. \ref{fig:RayTrace_sally_satra16}A.  If reflections from the bedrock are taken into account there will be additional solutions, which will be attenuated due to the longer paths.
\subsection{Attenuation length measurement}
A point-to-point attenuation length measurement using the hardware described here is on-going. The source of difficulties in this measurement includes the in-ice ray optics limiting the field of view of the antennas and the shadowing of the support cable adjacent to the receiving and transmitting antennas. In the upcoming season (2010-2011), a set of special antennas that are mounted symmetrically  around (rather than adjacent to) the support cable will be used for attenuation length measurements.

\subsection{Index of refraction measurement }
\begin{figure}[htbp]
\centering
\noindent
\includegraphics*[width=6cm]{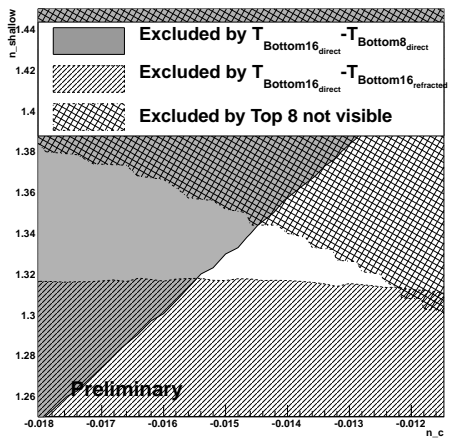}
\caption{Excluded parameters space for index of refraction based on time delays between hits . This area includes systematic uncertainties from the following uncertainties: timing resolution, slewing, geometry and $n_{deep}$. }
\label{fig:n_z}
\end{figure}

The extremely shallow location of the TDA units makes them sensitive to small variations in the index of refraction model, especially to $n_{shallow}$ and $n_{c}$.  Five out of six TDA units were able to trigger on a calibration pulse transmitted by the ACU  on ``Sally'' at $-245\,\mbox{m}$. The sixth unit that did not see the pulser (Top antenna at hole 8) was in the shaded area where no solution exists. 
Limits on the ice model parameters can be made by measuring the trigger time differences between 
different units as well as the time differences between the direct and refracted ray. This is 
illustrated in Fig \ref{fig:RayTrace_sally_satra16}B where an average WF from a pulser run for the 
top and bottom detectors in hole 16 is shown. The expected time delay between the direct and refracted ray for the top antenna was calculated to be about $14\,\mbox{ns}$, and the direct and refracted peaks are not resolvable. For the bottom antenna the simulated time delay was about $137\,\mbox{ns}$, in good agreement with the measurement. Figure \ref{fig:n_z} shows preliminary constraints on $n_{shallow}$ and $n_c$ based on combined time differences measured between detectors.

\begin{figure}[htbp]
\centering \noindent \includegraphics*[width=6cm]{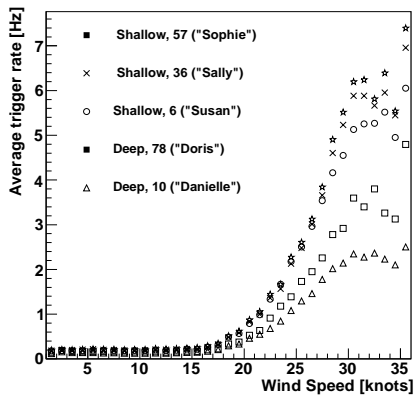}
\caption{Average trigger rate vs. wind speed. The meteorological data were taken from the ``Clean Air'' automatic weather station, operated by The Antarctic Meteorological Research Center (AMRC) and the Antarctic Automatic Weather Station (AWS) Program\cite{CAIR}.}
\label{fig:TriggerRateWind}
\end{figure}

\section{Wind-generated EMI}

Over four days in 2010 (July 15-16 and July 21-22), the volume of data collected by the radio detectors increased dramatically from $\sim200\,\mbox{MBytes}$ for a normal day to more than 5 GBytes.
The trigger rate study shows that the timing of the elevated noise rates coincided with the periods of winds stronger than $\sim20\,\mbox{knots}$ (Fig \ref{fig:TriggerRateWind}).  
Most of the WFs captured on the deep clusters during these strong winds were clearly down-going while the WFs on the shallow clusters are mostly saturated. 
Preliminary reconstruction study has pointed to an area around large structures in the Dark Sector 
as 
the origin of this interference. More sophisticated reconstruction efforts are in progress to better 
characterize the source of the noise.

Since the South Pole is electrically insulated electrostatic charge accumulates easily leading to a 
discharge causing EMI. There are two possible mechanisms as to how electrostatic charge builds up in 
strong winds: ``precipitation charging'' \cite{Dunham1966} where blowing snow causes structures to charge up, and ``snowstorm electrification'' \cite{GordonTaylor2009} where charge separation occurs near the snow surface.

\section{Conclusion}
Work is under way using the RF IceCube-radio-extension modules for characterizing the ice properties 
and the South Pole RF environment.
Understanding the nature and origin of the wind-related interference, index of refraction and attenuation lengths of ice will provide a valuable guide for designing a future GZK neutrino detector.  


%




\section*{Acknowledgments}
We acknowledge the support from the U.S. National Science Foundation-Office of Polar Program and U.S. National Science Foundation-Physics Division.

\bibliographystyle{model1-num-names}
\bibliography{mybibtex}

\begin{thebibliography}{13}
\expandafter\ifx\csname natexlab\endcsname\relax\def\natexlab#1{#1}\fi
\providecommand{\bibinfo}[2]{#2}
\ifx\xfnm\relax \def\xfnm[#1]{\unskip,\space#1}\fi
\bibitem[{Gusev and Zheleznykh(1983)}]{Gusev:1984gh}
\bibinfo{author}{G.~A. Gusev}, \bibinfo{author}{I.~M. Zheleznykh},
\newblock \bibinfo{title}{{Neutrino and Muon Detection from the radio emission
  of cascades created by them in natural dielectric media}},
\newblock \bibinfo{journal}{JETP Lett.} \bibinfo{volume}{38}
  (\bibinfo{year}{1983}) \bibinfo{pages}{611--614}.
\bibitem[{Kravchenko et~al.(2006)}]{Kravchenko:2006qc}
\bibinfo{author}{I.~Kravchenko}, et~al.,
\newblock \bibinfo{title}{{RICE limits on the diffuse ultra-high energy
  neutrino flux}},
\newblock \bibinfo{journal}{Phys. Rev.} \bibinfo{volume}{D73}
  (\bibinfo{year}{2006}) \bibinfo{pages}{082002}.
\bibitem[{Gorham et~al.(2010)}]{Gorham:2010kv}
\bibinfo{author}{P.~W. Gorham}, et~al.,
\newblock \bibinfo{title}{{Observational Constraints on the Ultra-high Energy
  Cosmic Neutrino Flux from the Second Flight of the ANITA Experiment}},
\newblock \bibinfo{journal}{Phys. Rev.} \bibinfo{volume}{D82}
  (\bibinfo{year}{2010}) \bibinfo{pages}{022004}.
\bibitem[{Gerhardt et~al.(2010)}]{Gerhardt:2010js}
\bibinfo{author}{L.~Gerhardt}, et~al.,
\newblock \bibinfo{title}{{A prototype station for ARIANNA: a detector for
  cosmic neutrinos}},
\newblock \bibinfo{journal}{arXiv:1005.5193}  (\bibinfo{year}{2010}).
\bibitem[{Hoffman(2010)}]{ara}
\bibinfo{author}{K.~Hoffman},
\newblock \bibinfo{title}{{The Askaryan Radio Array}},
\newblock \bibinfo{journal}{These Proceedings}  (\bibinfo{year}{2010}).
\bibitem[{Varner et~al.(2007)}]{labrador}
\bibinfo{author}{G.~S. Varner}, et~al.,
\newblock \bibinfo{title}{{The large analog bandwidth recorder and digitizer
  with ordered readout (LABRADOR) ASIC}},
\newblock \bibinfo{journal}{Nucl. Instrum. Meth.} \bibinfo{volume}{A583}
  (\bibinfo{year}{2007}) \bibinfo{pages}{447--460}.
\bibitem[{Landsman et~al.(2009)Landsman, Ruckman, and Varner}]{Landsman:2008rx}
\bibinfo{author}{H.~Landsman}, \bibinfo{author}{L.~Ruckman},
  \bibinfo{author}{G.~S. Varner},
\newblock \bibinfo{title}{{AURA - A radio frequency extension to IceCube}},
\newblock \bibinfo{journal}{Nucl. Instrum. Meth.} \bibinfo{volume}{A604}
  (\bibinfo{year}{2009}) \bibinfo{pages}{S70--S75}.
\bibitem[{Abbasi et~al.(2010)}]{IceCube}
\bibinfo{author}{R.~Abbasi}, et~al.,
\newblock \bibinfo{title}{{IceCube Collaboration Contributions to the 2009
  International Cosmic Ray Conference}},
\newblock \bibinfo{journal}{arXiv:1004.2093}  (\bibinfo{year}{2010}).
\bibitem[{Barwick et~al.(2005)}]{ice-atten}
\bibinfo{author}{S.~Barwick}, et~al.,
\newblock \bibinfo{title}{{South Polar in situ radio-frequency ice
  attenuation}},
\newblock \bibinfo{journal}{J. Glaciology} \bibinfo{volume}{51}
  (\bibinfo{year}{2005}) \bibinfo{pages}{231}.
\bibitem[{Kravchenko et~al.(2004)}]{ice-n}
\bibinfo{author}{I.~Kravchenko}, et~al.,
\newblock \bibinfo{title}{{South Polar in situ radio-frequency ice
  attenuation}},
\newblock \bibinfo{journal}{J. Glaciology} \bibinfo{volume}{50}
  (\bibinfo{year}{2004}) \bibinfo{pages}{522}.
\bibitem[{AMRC and AWC(2010)}]{CAIR}
\bibinfo{author}{AMRC}, \bibinfo{author}{AWC},
\newblock \bibinfo{title}{Surface observation from the cair aws},
\newblock
  \bibinfo{journal}{$ftp://amrc.ssec.wisc.edu/pub/southpole/surface_observatio%
ns$}  (\bibinfo{year}{2010}).
\bibitem[{Dunham(1966)}]{Dunham1966}
\bibinfo{author}{S.~B. Dunham},
\newblock \bibinfo{title}{Electrostatic charging by solid precipitation},
\newblock \bibinfo{journal}{J. Atmospheric Sciences} \bibinfo{volume}{23}
  (\bibinfo{year}{1966}) \bibinfo{pages}{412}.
\bibitem[{Gordon and Taylor(2009)}]{GordonTaylor2009}
\bibinfo{author}{M.~Gordon}, \bibinfo{author}{P.~Taylor},
\newblock \bibinfo{title}{The electric field during blowing snow events},
\newblock \bibinfo{journal}{Boundary-Layer Meteorology} \bibinfo{volume}{130}
  (\bibinfo{year}{2009}) \bibinfo{pages}{97--115}.
  \bibinfo{note}{10.1007/s10546-008-9333-7}.

\end{thebibliography}







\end{document}